\documentclass[a4paper,11pt]{article}
\pdfoutput=1 

\usepackage{graphicx}
\usepackage{epstopdf}
\usepackage{amsmath}
\usepackage{color}
\usepackage{icomma}
\usepackage{jcappub}
\usepackage{orcidlink}

\newcommand{\meter}{\textrm{m}}
\newcommand{\km}{\textrm{km}}

\newcommand{\Kelv}{\textrm{K}}

\newcommand{\GHz}{\textrm{GHz}}

\newcommand{\mJy}{\textrm{mJy}}

\newcommand{\Msun}{\textrm{M}_{\odot}}
\newcommand{\Lsun}{\textrm{L}_{\odot}}

\newcommand{\kpc}{\textrm{kpc}}
\newcommand{\Mpc}{\textrm{Mpc}}
\newcommand{\Gpc}{\textrm{Gpc}}

\newcommand{\kms}{\textrm{km}\,\textrm{s}^{-1}}

\newcommand{\MBH}{M_{\rm BH}}
\newcommand{\MBHRate}{\dot{M}_{\rm BH}}
\newcommand{\RS}{R_S}
\newcommand{\AProjBH}{A^{\rm proj}_{\rm BH}}

\newcommand{\OmegaLZ}{\Omega_{\Lambda,0}}
\newcommand{\OmegaMZ}{\Omega_{M,0}}
\newcommand{\OmegaRZ}{\Omega_{R,0}}
\newcommand{\OmegaBH}{\Omega_{\rm BH}}
\newcommand{\TCMB}{T_{\rm CMB}}
\newcommand{\TCMBZ}{T_{\rm CMB,0}}
\newcommand{\zeq}{z_{\rm eq}}
\newcommand{\DA}{D_A}

\newcommand{\sigmaSB}{\sigma_{\rm SB}}
\newcommand{\sigmaT}{\sigma_T}

\newcommand{\editOne}[1]{{#1}}
\newcommand{\mathendash}{\text{--}}

\title{Shadows of Giants: Constraints on Stupendously Large Black Holes from Negative Sources against the Cosmic Microwave Background}

\author[a]{Brian C. Lacki\,\orcidlink{0000-0003-1515-4857}}
\emailAdd{astrobrianlacki@gmail.com}
\affiliation[a]{Breakthrough Listen, Department of Physics, Denys Wilkinson Building, Keble Road, Oxford OX1 3RH, UK }

\abstract{Stupendously large astrophysical black holes (SLABs) are hypothetical black holes with masses of more than a trillion Suns. Because observable consequences of their existence have only recently been seriously considered, there have been relatively few constraints on their abundance. This work motivates a simple yet powerful constraint on SLABs: their huge shadows are visible against the cosmic microwave background. SLABs could thus appear as negative sources in microwave data. In fact, the shadow of a SLAB with a fixed mass becomes easier to detect with increasing redshifts past $1.6$ \editOne{where the angular diameter distance stars falling}. The limits are powerful enough to rule out SLABs of mass $\gtrsim 10^{17}\ \Msun$ within the last scattering surface, and imply $\OmegaBH \lesssim 10^{-5}$ for masses $10^{15} \mathendash 10^{18}\ \Msun$. I also discuss the effects of accretion and their implications for the limits: SLAB growth, positive accretion luminosity, and obscuring material.}

\begin{document}
\maketitle
\flushbottom

\section{Introduction}

Of the lines of evidence for black holes, two of the most direct have emerged in the past ten years: the detection of gravitational waves from merging black holes \citep{Abbott16,Abbott23-GWTC3} and the interferometric imaging of black hole shadows \citep{EHT19,EHT22}. The former provides a probe of spacetime near the event horizon, a largely clean window unaffected by the details of \editOne{any surrounding accretion flows}. At present, clear detections of gravitational waves from merging black holes come from LIGO and Virgo, sensitive to black hole masses in the stellar range \citep{Abbott16}. A stochastic background of nanohertz gravitational waves inferred with pulsar timing arrays may \editOne{originate from} binary supermassive black holes still with separations of thousands of AU \citep{Agazie23-GW,Agazie23-SMBH}. Black hole shadow imaging uses an accretion disk as a backlight \citep{Luminet79,Falcke00}. Only two black hole shadows have been observed, Sgr A* and M87*, both supermassive black holes in galactic centers \citep{EHT19,EHT22}. Together, then, each line of evidence \editOne{primarily} addresses one of the two main astrophysical classes of black holes: stellar or supermassive. The maximum mass of a supermassive black hole formed through conventional accretion is thought to be of order $\sim 10^{11}\ \Msun$ \citep{King16,Inayoshi16}, and the known ``ultramassive'' black holes are estimated to have masses of a few tens of billion Solar masses \citep[e.g.,][]{HlavacekLarrondo12,MeloCarneiro25}.

Could there be other black holes out there, though? Searches continue for intermediate mass black holes ($100 \mathendash 10^5\ \Msun$), which could represent the seeds of supermassive black holes or perhaps form in the hearts of star clusters \citep{Greene20}. But another hypothetical class has garnered attention in recent years: the primordial black holes (PBHs). PBHs would have formed shortly after the Big Bang from highly overdense perturbations. These perturbations may result from phase transitions in the early universe; the epoch of formation is related to the mass. Hypothesized PBHs have a full sweep of mass, from asteroidal to hypermassive \citep{Carr20}.\footnote{PBHs smaller than $10^{15}\ \mathrm{g}$ may have once existed but are thought to have evaporated by the present \citep{Carr10}.}

The resurgent interest in PBHs is motivated by several factors \citep{Carr24}. The failure to detect weakly interacting massive particles, either by Earthbound experiments or annihilation radiation in the cosmos, has led to greater interest in alternative dark matter candidates, including PBHs \citep{Bertone18}. The observation of galaxies at very high redshifts could indicate pre-existing massive seeds \citep[e.g.,][]{Yuan24,Gouttenoire24}. There have also been suggestions of evidence for new populations of black holes: some microlensing events appear to indicate Earth-mass objects that might be PBHs \citep{Niikura19}, and the LIGO black hole mergers have been attributed to a population of PBHs \citep{Bird16}. Constraints on PBHs at different masses are derived from many different astrophysical observables, including searches for black hole evaporation, microlensing, dynamical effects, annihilating dark matter haloes, and cosmic microwave background (CMB) distortions \citep{Carr21-Limits}. 

Carr et al. \citep{Carr21} proposed that some PBHs may be far greater in mass than even the supermassive black holes. These stupendously large black holes (SLABs) would have masses above a trillion suns -- greater than entire galaxies. Of course, such black holes could not explain the dark matter observed within galaxies, or even clusters, but conceivably they could exist in low numbers across the Universe. \editOne{Carr et al. questioned whether these SLABs are actually ruled out.} The limits set in ref. \citep{Carr21} were based on \editOne{their dynamical effects and} the energy released as they accrete matter in the early universe, which would distort the CMB, and the possible high-energy signatures from the halos of conventional dark matter they would acquire. They also note that the way they would lens the CMB would be detectable at $z \sim 0.5 \mathendash 10$; at the low end of the SLAB mass range, they would disrupt galaxies and seed large-scale structures too early. Deng \citep{Deng21} also considered CMB distortions from the collapse of possible cavities initially surrounding the PBH. Gerlach et al. \citep{Gerlach25} argued that SLABs would produce unacceptably large isocurvature perturbations in the CMB. Work by De Luca et al. \editOne{\citep{DeLuca26} suggests that the PBH formation would produce a detectable gravitational wave background.} Such black holes are evidently rare.

Stupendously large black holes were earlier considered in a very different context, however, as a hypothetical engineering project \citep{Lacki16}. These might be used as heat sinks to draw power from the CMB \citep{Opatrny17} or as gathering points \citep{Kardashev85,Smart12}; perhaps they might be created by massive transport of stars over intergalactic distances \citep{Hooper18}. Ref. \citep{Lacki16} noted that black holes of masses $\gtrsim 10^{15}\ \Msun$ would have enormous shadows. Even in empty intergalactic space, the shadows would blot out the CMB behind them. They would thus appear as cold spots, essentially negative flux sources. All unobscured SLABs have a shadow, even those formed long after recombination. 

\editOne{A much earlier still proposal to look for the shadow of a SLAB is found in ref. \citep{Paczynski86}, which considered the idea that a double quasar, QSO B1146+111 BC, is an illusion created by the lensing of a large black hole \citep{Turner86}. Paczy\'nski suggested that if this purported lensing SLAB had a mass of a quadrillion suns, its shadow against the CMB might be visible in the radio. We now have powerful experiments looking for anisotropies in the CMB near its spectral peak over the entire sky, enabling searches for these shadows.}

In this work, I consider the visibility of stupendously large black hole shadows, particularly primordial ones. In the absence of accretion, these shadows could yield very strong constraints. After all, a SLAB has an \emph{enormous} Schwarzschild radius, larger than entire galaxies at the uppermost end of the mass range:
\begin{equation}
\RS = \frac{2 G \MBH}{c^2} = 9.6\ \kpc \left(\frac{\MBH}{10^{17}\ \Msun}\right) .
\end{equation}
As ref. \citep{Lacki16} noted, a galaxy-sized blackbody with order unity temperature difference from the CMB is visible even at cosmological distances. Two features of the standard cosmology greatly aid this method. First, much of the comoving volume within the horizon is at relatively high redshift, $z \sim 20$. Second, the angular diameter distance of $\Lambda$CDM cosmology actually \emph{decreases} with redshifts beyond $z \sim 1.6$. Thus, SLAB shadows become \emph{easier} to detect at high redshift (as noted in refs. \citep{Bisnovatyi-Kogan18,Li20} for future observations of ``mere'' supermassive black holes). The last scattering surface has an angular diameter distance of only $\sim 13\ \Mpc$. At that distance, a $2 \times 10^{17}\ \Msun$ black hole would have a shadow roughly the angular size of the full Moon. Obviously, nothing like this is seen. Of course, SLABs may be accreting, particularly if their perturbation to the density field is not compensated by a surrounding void (as in \citep{Deng21}). The emission from such accretion flows could easily overcompensate the lost flux from the shadow, but then they should be even easier to detect as positive flux sources. The shadows thus set conservative constraints on SLABs that are nonetheless very powerful.

I assume a monochromatic SLAB mass distribution when deriving constraints. When specific results are needed, I adopt the flat \emph{Planck} 2018 cosmology, with $H_0 = 67.66\ \kms\,\Mpc^{-1}$, $\OmegaMZ = 0.3111$, recombination at $z = 1089.8$, and radiation-matter equality at $3387$ \citep{Planck20}. The present-day Hubble \editOne{parameter} $H_0$ is \editOne{expressed in terms of} $h_{70} \equiv H_0 / (70\,\km\,\sec^{-1}\,\Mpc^{-1})$. A zero subscript on a cosmological quantity denotes its value at $z = 0$.

\section{Naive estimates of black hole shadow visibility}
\subsection{The area and negative luminosity of SLABs}

Black holes famously trap light that enters their event horizons. That light does not have to be aimed directly into the horizon to be captured. It is thus meaningful to consider the black hole to be an absorber with a cross section for light extinction, which is
\begin{equation}
\AProjBH = 27 \pi \left(\frac{G \MBH}{c^2}\right)^2 = 0.0019\ \Mpc^2\ \left(\frac{\MBH}{10^{17}\ \Msun}\right)^2
\end{equation}
for a Schwarzschild (non-rotating) black hole \citep{Luminet79,Gralla19}. While rotating (Kerr) black holes can have distorted shapes, the cross section itself has a weak dependence on spin, reaching a minimum of $\sim 23 \pi (G \MBH/c^2)^2$ for extremal black holes viewed along the spin axis \citep{Leite18}. Large black holes have negligible temperature and extinguish background radiation of all \editOne{relevant} frequencies equally. Even when the black hole itself is unresolved, the capture and absorption of background light will appear as a brightness deficit, absent surrounding accreting matter.

There is an omnipresent background of radiation lying behind all unobscured SLABs since the surface of last scattering, the cosmic microwave background. While sources like active galactic nuclei and dust clouds add their own luminosity on top of the CMB, appearing as bright point sources in microwave sky maps \citep{Planck16}, a black hole shadow causes the sightline to appear darker than empty space. We can therefore consider the black hole to have an effective negative luminosity \editOne{(cf. \citep{Paczynski86})}. From the Stefan-Boltzmann law, and the fact that the absorbing surface area is four times the projected cross section,\footnote{\editOne{Equation 9 of \citep{Paczynski86} appears to be missing this factor of four.}}
\begin{equation}
\label{eqn:LShadow}
L_{\rm BH} = -\AProjBH \cdot 4 \sigmaSB \TCMB^4 = -108 \pi \frac{\sigmaSB \TCMB^4 G^2 \MBH^2}{c^4} = -6.1 \times 10^{10}\ (1 + z)^4\ \Lsun\,\left(\frac{\MBH}{10^{17}\ \Msun}\right)^2
\end{equation}
The CMB is hotter at high redshift, and thus, immensely brighter, even while isolated SLABs themselves remain almost perfectly dark. At the surface of last scattering, the negative luminosity of a black hole shadow on the CMB is over a trillion times greater than at present. As a result, SLABs become enormously antibrighter at high redshifts, more than compensating for their great luminosity distance. The shadows of SLABs are thus easier to detect at higher redshifts.

\subsection{The evolution of angular diameter distance}
\begin{figure}
\centerline{\includegraphics[width=9cm]{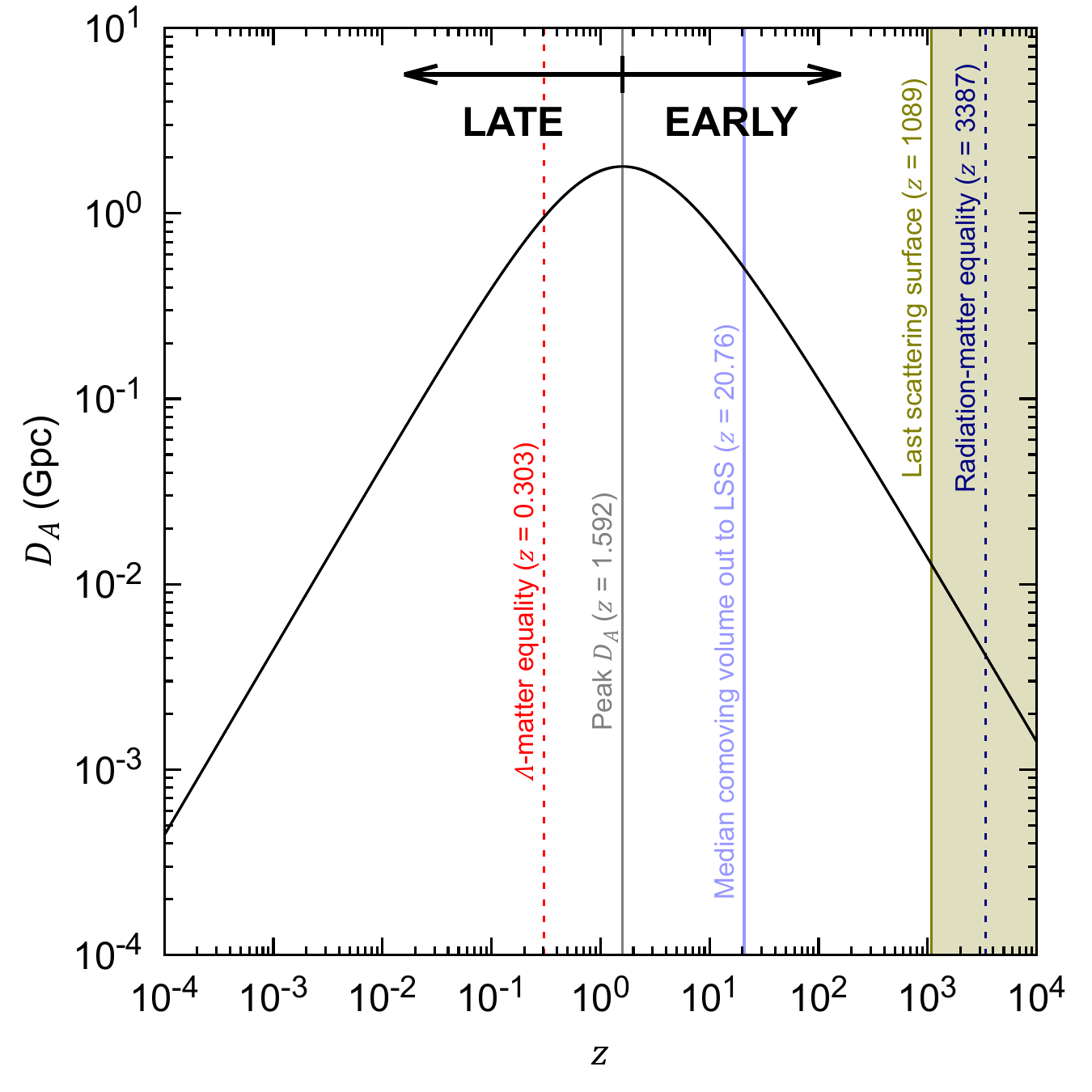}}
\caption{The evolution of angular diameter distance with redshift in the \emph{Planck} 2018 cosmology, showing how it falls past $z \sim 1.6$. \label{fig:DA}}
\end{figure}

The extreme negative luminosity of SLABs at high redshift implies large flux deficits, even though the CMB has been diluted and redshifted since those epochs. From our perspective, the angular diameter distance to the SLAB begins decreasing past a threshold redshift $z_A$. Thus, a SLAB of fixed mass blots out more and more of the CMB the higher the redshift it is. The angular diameter distance is defined as
\begin{equation}
\DA = D_M/(1 + z)
\end{equation}
where $D_M$ is the transverse comoving distance, equivalent to comoving distance for $\Lambda$CDM cosmology \citep{Hogg99}. The angular diameter distance reaches a maximum of $1.79\ \Gpc$ at $z_A = 1.592$ in the \emph{Planck} 2018 cosmology \citep{Planck20}, decreasing to a mere $13\ \Mpc$ at the surface of last scattering (Figure~\ref{fig:DA}). We can divide the comoving volume within the horizon into a \emph{late volume} with $z < z_A$ and an \emph{early volume} with $z \ge z_A$. 

\begin{figure}
\centerline{\includegraphics[width=9cm]{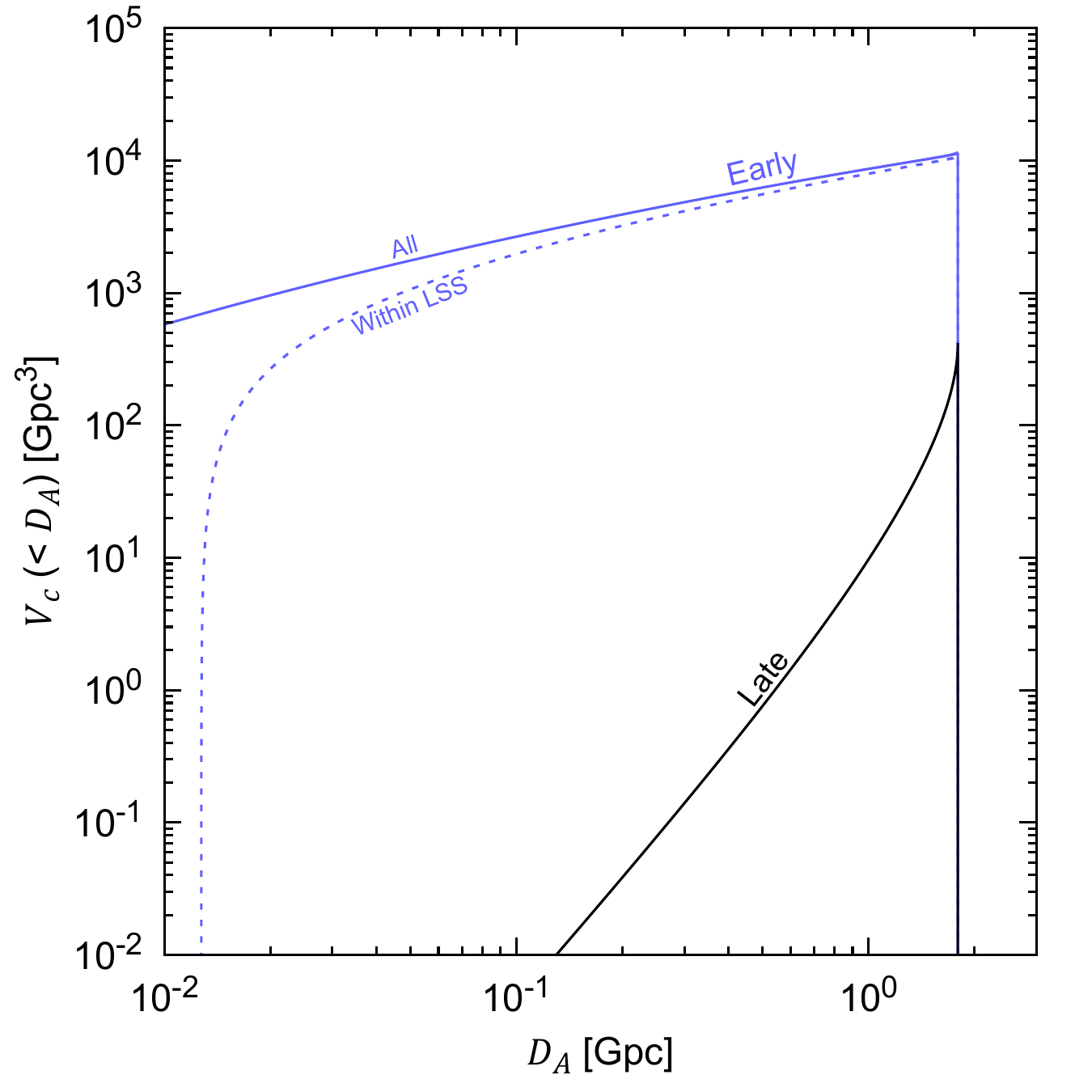}}
\caption{The cumulative comoving volume with an angular diameter distance below a given value. The black line shows the low-redshift late volume, where the angular diameter distance is small simply because the object is nearby. At high redshift, there is an ``early'' volume where angular diameter distance decreases well below a Gpc, shown in blue, with the solid line showing the total early volume to the Big Bang and the dashed line only including the early volume out to the last scattering surface. \label{fig:ComovingVolume}}
\end{figure}

As long as SLABs are sufficiently isolated that mergers are infrequent (cf. \citep{Atal21}), and assuming there are no late-time mechanisms for the creation of SLABs, their comoving number density should be the same at all redshifts, even if their masses evolve. The limits are thus derived from how much comoving volume we can rule out a SLAB in. Figure~\ref{fig:ComovingVolume} shows the amount of volume at each value of $\DA$. Within the late volume, $\DA$ increases with redshift, showing the Euclidean cubic dependence for $z \lesssim 1$. There is a limited amount of comoving volume when this dependency holds -- of course, we only expect $\sim 4\ \Gpc^3$ of comoving volume for comoving distances out to $1\ \Gpc$, for example. Beyond $z \sim 1$, the amount of comoving volume starts growing faster than cubic, reaching a maximum of $420\ \Gpc^3$ in the whole late volume. In contrast, the far early regime (blue) has $\DA$ decreases with redshift. The amount of early comoving volume is vast, $\sim 10^4\ \Gpc^3$. In fact, the median $\DA$ in terms of comoving volume is about half a Gpc, with over $10\%$ of the volume having $\DA \lesssim 100\ \Mpc$.  It is these vast volumes combined with the immense, easily detectable negative luminosity of early SLABs that allow us to set powerful limits.

\subsection{The detectability of SLAB shadows}
Microwave point surveys report a minimum spectral flux to which they are reasonably complete. I assume that a source with a negative flux of the same magnitude would also \editOne{have been reported. Their detection in CMB experiments is certainly possible; hot plasma in galaxy clusters is detected as a negative source at low frequencies due to the Sunyaev-Zeldovich effect, for example \citep{Birkinshaw99}.} The negative spectral flux of the SLAB shadow against the CMB is easily found from the Planck law for a blackbody and the shadow's projected area on the sky:
\begin{equation}
S_{\nu} = -\frac{2 h \nu^3}{c^2} \frac{1}{\exp(h \nu/(k_B \TCMBZ)) - 1} \frac{\AProjBH}{\DA^2} ,
\end{equation}
where the CMB temperature is its observed temperature at $z = 0$.  Microwave sky maps have relative poor sensitivity to point sources, so it takes a very large black hole to be visible against the CMB at extragalactic distances, even with the quadratic dependence on mass.

The allowed abundance of SLABs is directly related to the comoving volume where $\DA$ is small enough for the shadow to be detectable. If SLABs are distributed independently of each other, the distribution of the number of black holes is Poissonian. A null result lets us set an upper limit of $\bar{N}$ SLABs of mass $\MBH$ in a survey volume $V_c^{\rm obs}(\MBH)$ where the shadow is large enough to be detectable and within the survey footprint; for a Poissonian distribution, the 95\% confidence upper limit is given by $\bar{N} = 3.00$ \citep{Gehrels86}. Then the maximum allowed number density of SLABs of mass $\MBH$ is $n^{\rm max}_{\rm SLAB} = \bar{N}/V_c^{\rm obs}(\MBH)$.

It is common to phrase constraints on PBHs in terms of the critical comoving density of the Universe, which is $\rho_c = 3 H_0^2 / (8 \pi G)$. Assuming the SLABs have a constant, unchanging mass of $\MBH$, we find:
\begin{equation}
\OmegaBH(\MBH) \le \frac{8 \pi \bar{N} G \MBH}{3 H_0^2 V_c^{\rm obs}(\MBH)} = 0.0022\ h_{70}^{-2} \left(\frac{\MBH}{10^{17}\ \Msun}\right)\left(\frac{V_c^{\rm obs}(\MBH)}{\Gpc^3}\right) \left(\frac{\bar{N}}{3}\right) .
\end{equation}

\subsection{Formation epoch of primordial SLABs}
The initial mass of a PBH that forms from an overdense perturbation collapsing during the radiation epoch is thought to be of order the mass-energy in the horizon \citep{Carr21-Limits}:
\begin{equation}
\MBH(z_{\rm form}) \sim \frac{4\pi}{3} \left(\frac{c}{H(z_{\rm form})}\right)^3 \frac{3 H(z_{\rm form})^2}{8 \pi G} \sim \frac{1}{2} \frac{c^3}{G H(z_{\rm form})} .
\end{equation}
THe Hubble factor at this redshift is given by $H(z_{\rm form}) = H_0 [\OmegaLZ + \OmegaMZ (1 + z_{\rm form})^3 + \OmegaRZ (1+z_{\rm form})^4]$. The dark energy term is \editOne{assumed to be} negligible in this era. We can approximate the radiation density by comparing it to the matter density at the redshift of matter-radiation equality ($\zeq \approx 3400$): $\OmegaRZ = \OmegaMZ / (1 + \zeq)$. We then find:
\begin{align}
\nonumber \MBH(z_{\rm form}) & \sim \frac{1}{2} \frac{c^3}{G H_0 \sqrt{\OmegaMZ}} \frac{(1 + \zeq)^{1/2}}{(1 + z)^2} \left(1 + \frac{1 + \zeq}{1 + z}\right)^{-1/2} \\
                              & \sim 4 \times 10^{17}\ \Msun\ \left(\frac{1 + z}{1 + \zeq}\right)^{-2} h_{70}^{-1} \left(\frac{\OmegaMZ}{0.3}\right)^{-1/2} \left(\frac{1 + \zeq}{3400}\right)^{-3/2} .
\end{align}
If this is how SLABs formed, then \editOne{any PBHs} with masses up to $\sim 10^{17}\ \Msun$ formed before decoupling and the entire volume out to the last scattering surface may then be used for constraints. If there are SLABs with $\MBH \gg 10^{17}\ \Msun$, then the comoving volume in which they may be found is reduced. Note, however, that the shadow constraint would be so powerful as to eliminate the possibility of any SLAB that massive in the entire comoving volume: they would blot out entire square arcminutes of the CMB even at the maximum angular diameter distance of $2\ \Gpc$ -- indeed, the shadows alone are large enough to be resolved both by \emph{Planck}'s high frequency detectors \emph{and the naked eye}, although invisible to the latter.

\section{How accretion affects the shadow limits}

The calculations have assumed that the population of SLABs is static, nonluminous, and unobscured. But in reality, SLABs could interact with the background matter and fields of the Universe at large. The previous assumptions could still be appropriate if the SLABs are born from compensated fluctuations. These have a net mass of zero -- the central overdensity that collapses into the PBH is surrounded by a void of equal mass deficit \citep{Deng21}. When a compensated fluctuation is spherically symmetrical, the PBH and void cancel each other out by Birkhoff's theorem, and the external Hubble flow is unaffected as long as pressure forces can be neglected. As a result, accretion may be ignored in these compensated scenarios.

But it is also possible that the initial fluctuation is uncompensated, so that the excess mass of the SLAB on the background serves to locally perturb the Hubble flow. In the matter-dominated epoch, the SLAB may accrete both baryonic and dark matter. Primordial SLABs would be so large that their accretion would be limited by the cosmological expansion; estimates based on Bondi accretion do not apply \citep{Carr21}. In the Bertschinger (B85) self-similar model of infall onto a perturbation, the surrounding matter is treated as a series of initially expanding concentric shells surrounding the collapsed perturbation \citep{Bertschinger85}. Since the perturbation is spherically symmetric, each shell essentially has Keplerian dynamics. The excess mass causes each shell to decelerate, eventually reaching apocenter at its turnaround radius in half an orbital period. It then falls inward, reaching pericenter after nearly a full orbital period. As a result, the SLAB is surrounding by a self-similar infalling halo of matter, transitioning from a $\rho \propto r^{-9/4}$ to $\rho \propto r^{-3/2}$ density profile in its core as the black hole comes to dominate the mass \citep{Bertschinger85}.

What happens when the infalling matter reaches the center of the halo depends on various considerations. In the simplest case, the matter has \editOne{no} angular momentum and is collisionless. Then it falls directly into the black hole, adding to its mass and causing it to grow \citep{Mack07}. While this may apply to (presumably) collisionless dark matter, large enough peculiar velocities result in pericenters outside the event horizon. As a result, the dark matter forms a growing halo around the black hole, while the SLAB itself may have little growth \citep[cf.][]{Ricotti08}.

\subsection{Growth of the SLAB from direct infall}

\editOne{I shall assume the infalling matter has no angular momentum and thus accretes onto the SLAB.} According to B85, during the matter-dominated epoch, the mass within the turnaround radius is 
\begin{equation}
M_{\rm ta}(t) \propto \MBH(t) \cdot (t/\bar{t})^{2/3} \propto \MBH(z) \left(\frac{1 + z}{1 + \bar{z}}\right)^{-1}
\end{equation}
where $\bar{t}$ is the age of the Universe at a reference redshift of $\bar{z}$. Now, matter that is reaching pericenter at time $t$ was at its turnaround radius about half an orbital period ago (neglecting a short early interval until the steady state evolution begins), at time $t/2$. All of this matter has been ingested into the black hole, so the growth of the SLAB mass is
\begin{equation}
\label{eqn:MBHGrowthSS}
\MBH(z) = \MBH(\bar{z}) \left(\frac{1 + z}{1 + \bar{z}}\right)^{-1}
\end{equation}
during the matter-dominated epoch \citep{Mack07}. In general, for $t \gg \bar{t}$, the accreted mass is far greater than the seed PBH mass in this scenario. If all of the infalling matter is non-radiating and transparent, then the relevant effects of accretion are that the SLABs have much smaller shadows at high redshift. This negates the advantage of small $\DA$ at high redshift; although that distance measure is decreasing, the shadow's effective area shrinks faster. 

As dark energy dominates the cosmic expansion, the accretion flows are choked off \citep{Subramanian00}. In the standard $\Lambda$CDM cosmology, the expected final black hole mass is around 40\% the naive matter-dominated expectation \citep{Mack07}. Additionally, because the Keplerian period of matter still infalling in the SLAB's halo is comparable to the Hubble time, some of the last mass shells may still be infalling. However, I ignore this factor of $\sim 2 \mathendash 3$ correction\editOne{; the epoch of dark energy dominance only makes up a small fraction of the Universe's observable comoving volume.}

\subsection{The scale of the accretion flow}
\label{sec:AccretionFlow}
Baryonic matter accreting onto a SLAB may release tremendous luminosity as it falls in, or obscure its shadow entirely. For all but the largest SLABs at the highest redshifts (lowest $\DA$), which may actually be resolved, the accretion flow will be blended with the shadow, potentially dominating the net flux observed.

The core of the accretion flow is the most relevant for the in the B85 model: most of the accretion luminosity and obscuration is concentrated near the black hole. The sheer scale of the flow is worth noting, as it may yield other constraints. Matter at the turnaround radius $R_{\rm ta}(z)$, beginning its fall into the black hole, is at apocenter, one-half of an orbital period from its accretion. During the matter epoch, the mass within this region is $M_{\rm ta}(z) \approx 2^{2/3} \MBH(z)$. From Kepler's Third Law, the proper radius of the infalling region
\begin{equation}
R_{\rm ta}^{\prime}(t) \approx \left[\frac{2^{8/3}}{9\pi} \frac{G \MBH(z)}{H_0^2 \OmegaMZ (1 + z)}\right]^{1/3} \approx 40\ \Mpc\,\left(\frac{\MBH(z)}{10^{17}\ \Msun}\right)^{1/3} h_{70}^{-2/3} \left(\frac{\OmegaMZ}{0.3}\right) (1 + z)^{-1} .
\end{equation}
An uncompensated primordial SLAB would thus warp the large-scale structure of the universe. For masses of a trillion Suns, the turnaround radius is nearly a megaparsec; for masses of a quintillion Suns, this infalling halo would be as large as a cosmic void. The infalling matter would have velocities of order $\sim \sqrt{G \MBH(z)/R_{\rm ta}^{\prime}}$, around $10^3\ \kms$ when $\MBH \sim 10^{15}\ \Msun$ and $10^4\ \kms$ for $\MBH \sim 10^{18}\ \Msun$.

\subsection{Accretion luminosity}
Suppose the black hole grows self-similarly as in the B85 model. The mass growth rate from the accreting matter is found by differentiating equation~\ref{eqn:MBHGrowthSS}. A fraction $\eta_{\rm ss}$ of the rest-energy of the infalling mass is converted into luminosity as it is ingested. This factor accounts for both the radiative efficiency of accretion and only $\sim 1/6$ of the matter in the Universe being baryonic. Thus $\eta_{\rm ss} \sim 0.01 \mathendash 0.1$ would be a typical value for radiatively efficient accretion, with much smaller values possible for \editOne{radiatively inefficient} accretion modes. In the matter-dominated epoch, the luminosity from this steady-state accretion mode is:
\begin{align}
\nonumber L_{\rm ss} & \approx \eta_{\rm ss} \MBHRate(z) c^2 = \eta_{\rm ss} \MBH(\bar{z}) c^2 H_0 \sqrt{\OmegaMZ} (1 + \bar{z}) (1 + z)^{1/2} \\ 
\label{eqn:LAccSteadyState}
                     & \approx 2.7 \times 10^{18}\ \Lsun\,h_{70}\left(\frac{\MBH(\bar{z})}{10^{17}\ \Msun}\right) \left(\frac{\eta_{\rm ss}}{0.01}\right) \left(\frac{\OmegaMZ}{0.3}\right)^{1/2} (1 + \bar{z}) \left(\frac{1 + z}{22}\right)^{1/2} ,
\end{align}
with the black hole mass normalized to its value at redshift $\bar{z}$. I have scaled to a redshift $z \sim 21$, within which about half the comoving volume out to the last scattering surface is contained. The luminosity decreases with time, even as the black hole mass itself increases. Larger black holes would, all else being equal, have brighter accretion flows.

\begin{figure}
\centerline{\includegraphics[width=9cm]{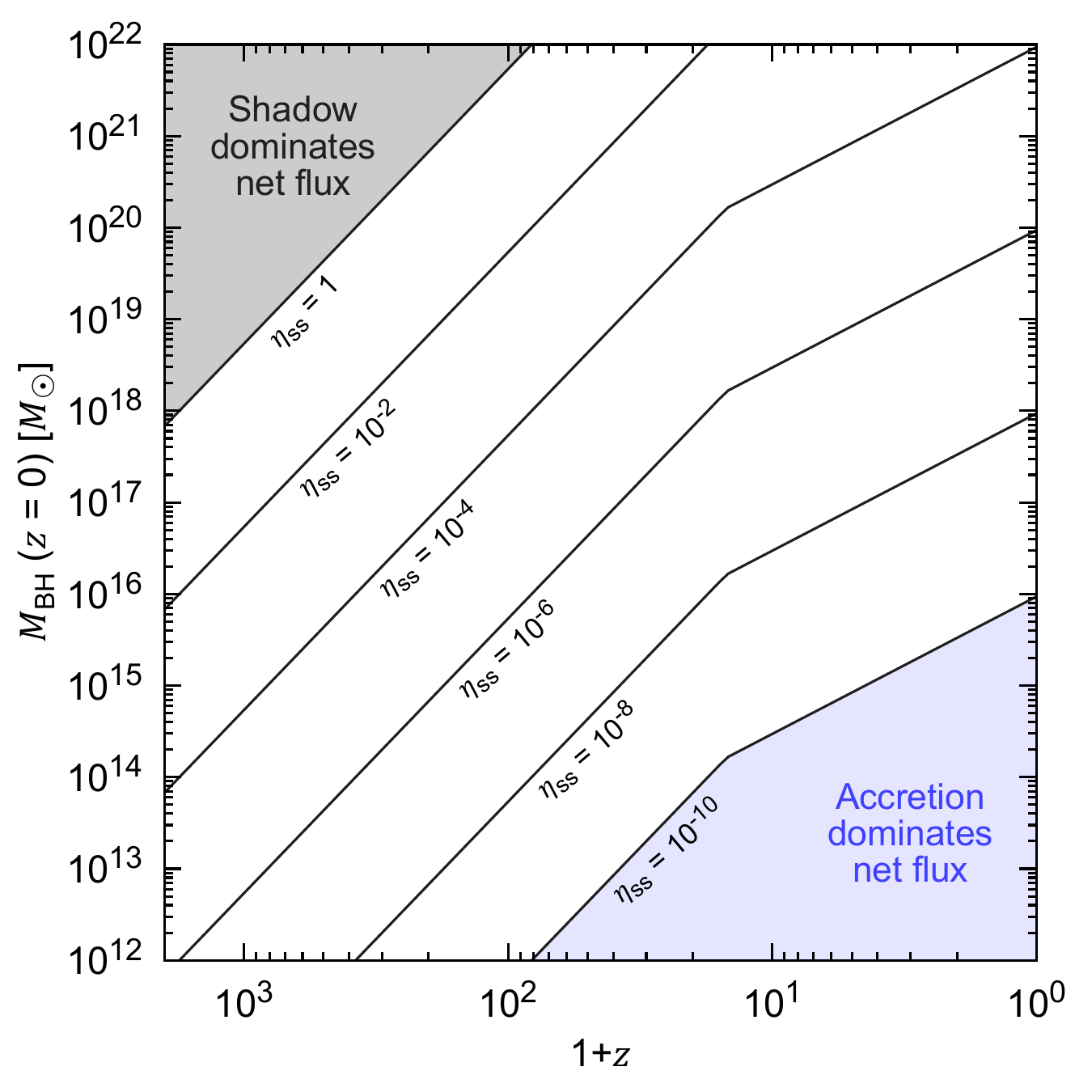}}
\caption{Chart of whether the luminosity deficit of the shadow or the luminosity from accretion dominates the net bolometric flux from SLABs of various masses, redshifts, and radiative efficiencies in the B85 steady-state accretion model. Black holes above a line have a net flux dominated by the shadow; those below it would have a net bolometric flux dominated by the accretion luminosity. Black holes are assumed to grow as $\MBH(z) \propto (1 + z)^{-1}$. There is a change in slope at later times as the baryonic accretion is no longer limited by the Eddington limit. Note the transition between shadow-dominated and accretion-dominated net flux will differ at individual frequencies.\label{fig:LuminosityRegimes}}
\end{figure}

We can compare this with the luminosity deficit resulting from the shadow on the CMB (equation~\ref{eqn:LShadow}). The accretion flow is brighter if
\begin{align}
\nonumber \MBH(\bar{z}) & < \frac{\eta_{\rm ss} c^6 H_0 \sqrt{\OmegaMZ}}{108 \pi \sigma_{\rm SB} \TCMBZ^4 G^2 (1 + \bar{z})(1 + z)^{3/2}} \\
                        & < 9.2 \times 10^{21}\ \Msun\,h_{70} \left(\frac{\eta_{\rm ss}}{0.01}\right) \left(\frac{\OmegaMZ}{0.3}\right)^{1/2} (1 + \bar{z})^{-1} \left(\frac{1 + z}{22}\right)^{-3/2}.
\end{align}
Note that this is for the net \emph{bolometric} flux; the net flux in a particular frequency range depends on the redshifted emission spectrum of the accretion flow, which could be very complicated. The shadow dominates the net emission for the largest black holes because the deficit grows quadratically with the mass (and Schwarzschild radius) while the mass accretion rate only grows linearly with mass. However, accretion is more dominant at later times for a fixed black hole seed mass: the energy density of the CMB decreases as the fourth power of redshift, shrinking the flux deficit rapidly (see Figure~\ref{fig:LuminosityRegimes}). 

The enormous luminosities potentially run afoul of the Eddington limit, in which the radiation pressure on ionized gas counterbalances gravity. Either the baryonic mass accretion rate is choked, \editOne{it} transitions to a radiatively inefficient mode, or assumptions like steady-state accretion or spherical symmetry that go into the Eddington limit \citep{Frank02} are strongly violated. For ionized gas, radiation pressure acts on Thomson scattering on free electrons, resulting in a luminosity limit of:
\begin{equation}
L_{\rm E}(z) = 4 \pi G \MBH(z) m_P c / \sigmaT, 
\end{equation}
where $\sigmaT = 6.65 \times 10^{-29}\ \meter^2$ is the Thomson cross scattering section \editOne{and $m_P$ is the mass of the proton}. This maximum luminosity still has a greater magnitude than the luminosity deficit of the shadow when:
\begin{equation}
\MBH(\bar{z}) \lesssim \frac{m_P c^5}{27 G \sigmaT \sigmaSB \TCMBZ^4} (1 + \bar{z})^{-1} (1 + z)^{-3} = 5.1 \times 10^{23}\ \Msun\,(1 + \bar{z})^{-1} \left(\frac{1 + z}{22}\right)^{-3}
\end{equation}

The Eddington limit thus acts to quench accretion luminosity in the early Universe, for redshifts 
\begin{equation}
z \gtrsim \left(\frac{4 \pi G m_P}{\eta_{\rm ss} \sqrt{\OmegaMZ} H_0 \sigmaT c}\right)^{2/3} = 320 \left(\frac{\eta_{\rm ss}}{0.01}\right)^{-2/3} \left(\frac{\OmegaMZ}{0.3}\right)^{-1/3} h_{70}^{-2/3} .
\end{equation}
Note that the radiation pressure would not affect collisionless dark matter, and thus the growth of the black hole itself would not necessarily be affected much.

Does the prospect of accretion luminosity render the shadow constraints moot? Not necessarily, because the limits ultimately derive from the detectability of point sources against the CMB, whether positive or negative flux. It is unlikely that the accretion luminosity excess just so happens to cancel out the excess luminosity. We expect that one dominates the other. Moreover, for a given black hole seed mass and radiative efficiency, that cancellation only holds for a single redshift; we would thus expect to see net negative sources at high redshift \emph{and} net positive sources at low redshift, and we have signs of neither. The accretion luminosity of the low redshift sources could make them the most luminous things in the Universe, and thus easily detected by now.

\subsection{Obscuration by the accretion flow}
Another potential problem is that the ionized material around the SLAB \editOne{could} be opaque, preventing its shadow from being visible. In the early Universe, only hydrogen and helium exists in the intergalactic medium, with the main source of opacity in the accreting halo being direct Thomson scattering from ionized infalling material. If we assume the accretion flow is in free-fall and spherically symmetric, the maximum Thomson optical depth around the shadow is
\begin{equation}
\tau_T(z) = \int_{\RS(z)}^{\infty} f_{\rm ion}(r) n_e(r) \sigmaT dr \approx \int_{\RS(z)}^{\infty} \frac{f_{\rm ion}(r) \dot{M}_b \sigmaT}{4 \pi m_H r^2 v(r)} dr
\end{equation}
for a baryonic mass accretion rate of $\dot{M}_b$ and ionization fraction $f_{\rm ion}$. When the matter is in free-fall, $v(r) = \sqrt{2 G \MBH/r}$ near the core of the halo. In the B85 model, 
\begin{equation}
\tau_T \approx \frac{f_b f_{\rm ion} \sqrt{\OmegaMZ} c H_0 \sigmaT}{4\pi m_H G} (1 + z)^{3/2} \approx 0.36 h_{70} \left(\frac{f_b f_{\rm ion}}{0.2}\right)\left(\frac{\OmegaMZ}{0.3}\right)^{1/2} \left(\frac{1 + z}{22}\right)^{3/2} .
\end{equation}
Almost all of the optical depth results from matter near the event horizon. In this model, obscuration becomes important for $z \gtrsim 40 f_{\rm ion}^{-2/3}$, regardless of black hole mass. Note, however, that the ionization fraction for these black holes may be low. A naive estimate of the temperature of an optically thick accretion flow can be found by modeling it as a blackbody with emitting area $\zeta \AProjBH$, giving a value of $\sim 6,800\ \Kelv (\MBH(z)/10^{12}\ \Msun)^{-1/4} (\eta_{\rm ss}/0.01)^{1/4} (\zeta/10)^{-1/4} ((1 + z)/22)^{3/8}$ for the (non-Eddington limited) steady-state accretion luminosity. According to this calculation, quadrillion solar mass SLABs have effective temperatures of at most $5,000\ \Kelv$ at the surface of last scattering, and a mere $1,200\ \Kelv$ at $z = 21$.

\subsection{Complications of baryonic accretion}
In practice, the structure of the accretion flow may differ greatly from the steady-state model considered above. Any initial peculiar velocities will prevent direct ingestion of the matter onto the black hole; instead it would form a baryonic core around the SLAB. While this would allow matter to accumulate, the scale of the resulting accretion disk would be much larger than the black hole and thus the mass would be more dilute. Additionally, the higher densities allow for more effective radiative cooling, potentially causing most of the gas to remain neutral. The gas may collapse to form stars (\editOne{which would be Population III for high redshift SLABs}), sealing away most of the mass and opening up lines of sight to the shadow. \editOne{Additionally, at late epochs, a fraction of the baryonic material would already be condensed into galaxies.} A more detailed study of accretion in these unusual conditions is necessary to fully account for its effects.

\section{Resultant limits on SLAB densities}
\editOne{\subsection{Limits from non-observation of SLAB shadows in CMB experiments}}
I use the \editOne{radio and} microwave point source surveys listed in Table~\ref{table:MicrowaveSourceData} to set upper limits on the abundance of SLABs, assuming the net flux is dominated by the shadows. The \emph{Planck} catalog of compact sources covers much to all of the sky \editOne{in the lower frequency bands} and thus can be used to rule out the largest SLABs \citep{Planck16}. \editOne{The in-progress VLASS survey at the VLA also covers most of the sky, reaching milliJansky flux densities, but is conducted at 3 GHz, where the Rayleigh-Jeans tail suppresses the shadow negative flux.} The other two use the more sensitive South Pole Telescope, which images smaller regions of the sky to greater depths \citep{Everett20,Archipley25}. This is the trade-off: a deeper survey requires more integration time and covers less of the sky, allowing it to be sensitive to smaller shadows in its field (smaller SLAB masses) but potentially missing the majority of shadows (allowing higher SLAB densities).

Two basic models of SLAB growth are considered: (1) the SLAB does not grow at all since the surface of last scattering and (2) self-similar growth with $\MBH(z) = \MBH(0) / (1 + z)$ as in the B85 evolution. The latter ignores the quenching of accretion by the accelerating expansion in our dark energy epoch, but as argued previously, this inhibition only occurs very recently.

\begin{table}
\centering
\begin{tabular}{cccccc}
Facility & Survey & Reference & $\nu$ & $\omega/(4\pi)$ & $F_{\nu}$ \\ 
 & & & ($\GHz$) & & ($\mJy$) \\
\hline
\emph{Planck} & PCCS2                   & \citep{Planck16} & 30  & 1.000  & 427 \\
              &                         &                  & 44  & 1.000  & 692 \\
							&                         &                  & 70  & 1.000  & 501 \\
							&                         &                  & 100 & 0.850  & 269 \\
							&                         &                  & 143 & 0.850  & 177 \\
							&                         &                  & 217 & 0.649  & 152 \\
							&                         &                  & 353 & 0.476  & 304 \\
							&                         &                  & 545 & 0.470  & 555 \\
							&                         &                  & 857 & 0.463  & 791 \\
\hline
\editOne{VLA} & \editOne{VLASS}         & \citep{Lacy20}   & \editOne{3.0} & \editOne{0.82} & \editOne{0.35} \\
\hline
SPT-SZ        & Everett et al. (2020)   & \citep{Everett20} & 95  & 0.0613 & 12.89 \\
	            &                         &                   & 150 & 0.0613 & 7.60 \\
							&                         &                   & 220 & 0.0613 & 26.83 \\
\hline
SPT-3G        & Archipley et al. (2025) & \citep{Archipley25} & 95  & 0.0014 & 1.7 \\
              &                         &                     & 150 & 0.0014 & 2.0 \\
							&                         &                     & 220 & 0.0014 & 6.5 \\
\end{tabular}
\caption{Limits on point sources in high-frequency microwaves from three surveys (PCCS2 referring to the Second \emph{Planck} Catalog of Compact Sources \editOne{and VLASS to the Very Large Array Sky Survey}). The central frequency of each band is $\nu$; $\omega$ is the solid angle covered by the survey in the listed band; $F_{\nu}$ is the flux density sensitivity for the band. The sensitivities are for 90\% completeness in PCCS2; the $5\sigma$ point source limit for \editOne{VLASS} and SPT-3G\editOne{; and} the median 95\% completeness over the fields observed in \citep{Everett20} for SPT-SZ (the worst case $F_{\nu}$ would be $14.13$, $8.05$, and $29.70\ \mJy$ for $90$, $150$, and $220\ \GHz$, respectively). \label{table:MicrowaveSourceData}}
\end{table}

\begin{figure*}
\centerline{\includegraphics[width=8cm]{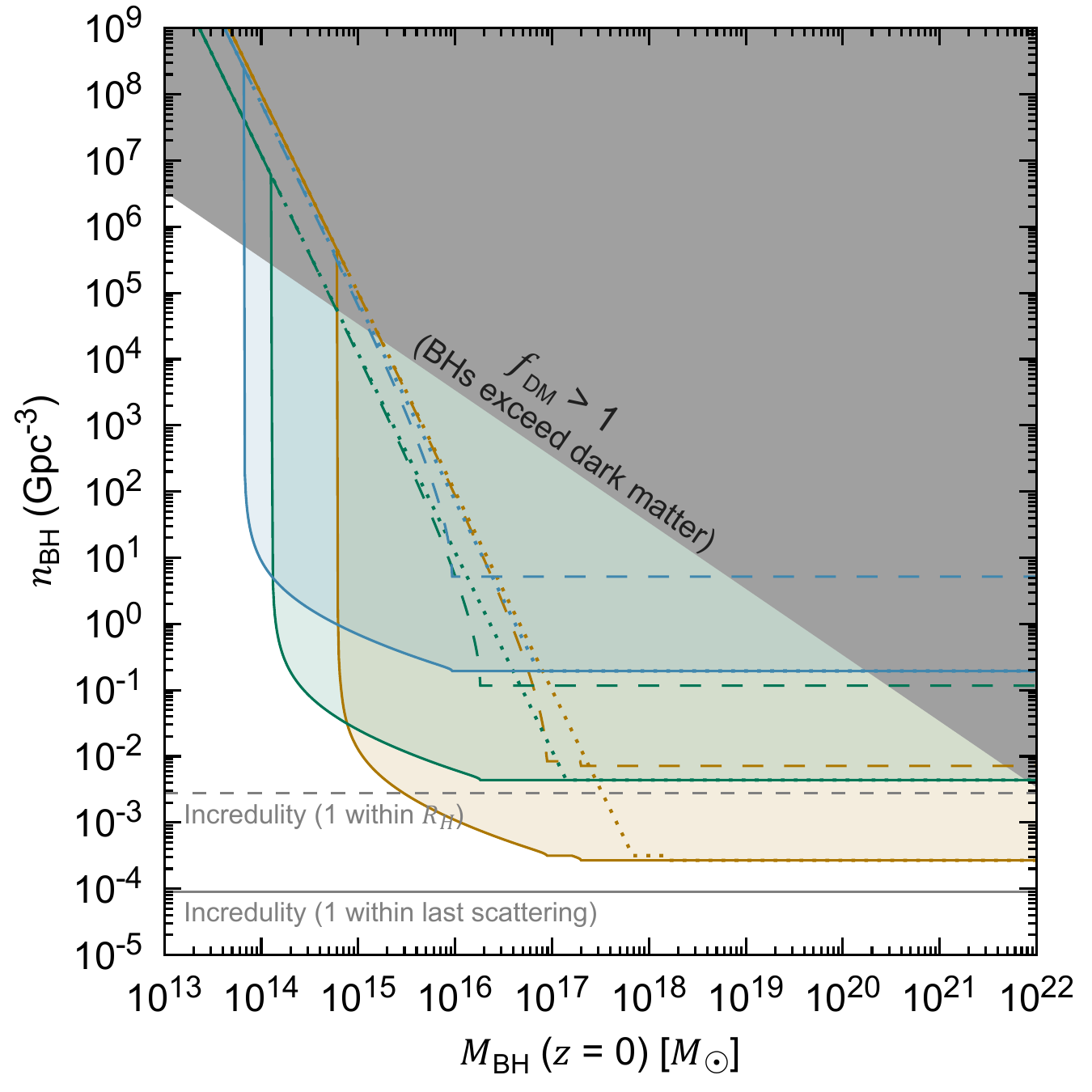}\includegraphics[width=8cm]{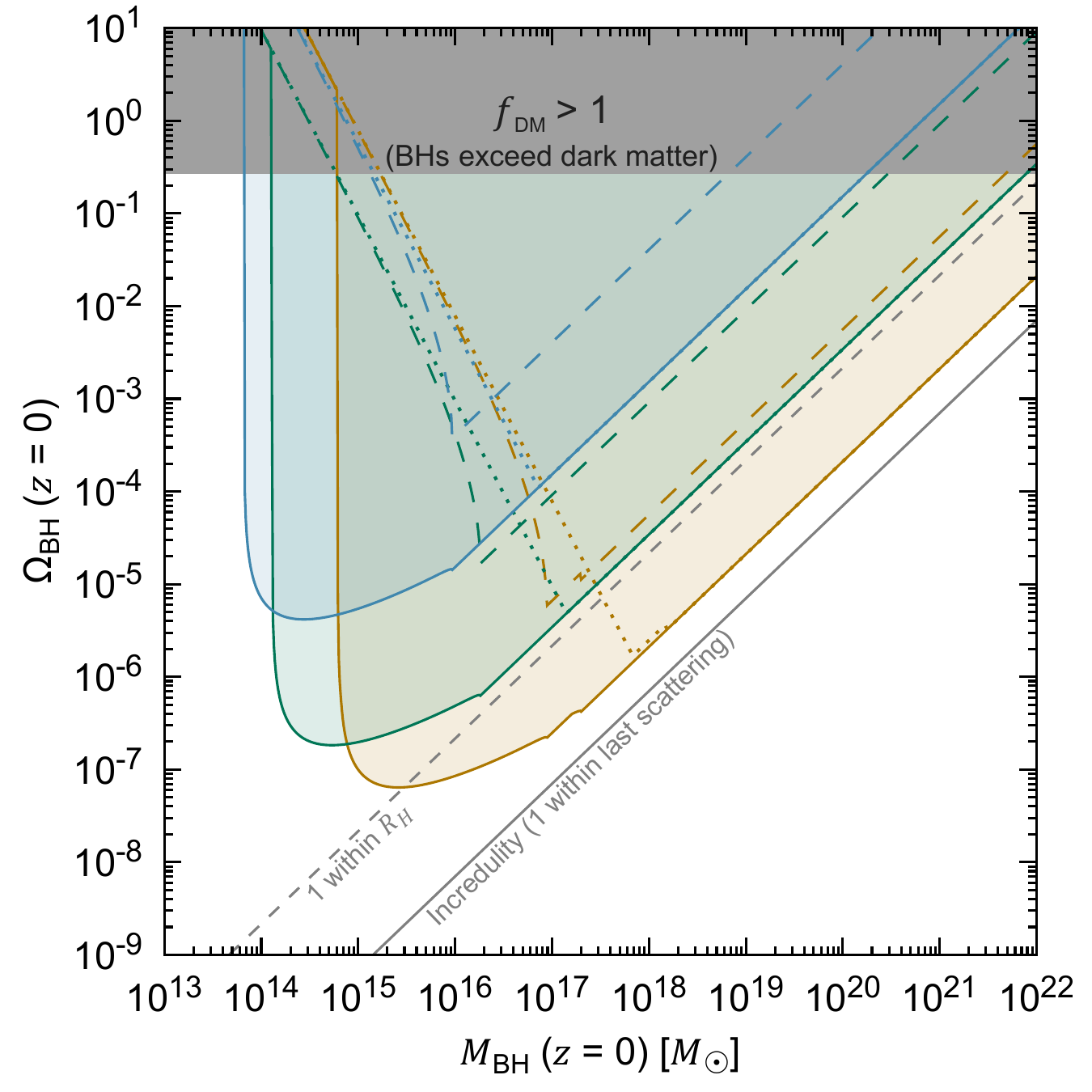}}
\caption{Limits on SLABs from black hole shadows alone (95\% confidence upper limits), ignoring the possible luminosity and obscuration from baryonic accretion flows. These are in terms of comoving number density (left) and fraction of the cosmological critical density (right). The solid lines represent the limits including both the early and late volume for SLABs of constant mass. The dashed lines represent the limits when only including the late comoving volumes. The dotted lines show the limits for black holes growing as $\MBH \propto (1 + z)^{-1}$. Three surveys are plotted: the \emph{Planck} Catalog of Compact Sources 2 (orange), SPT-SZ (green), and SPT-3G (blue). \editOne{The VLASS limits are not shown but are nearly identical to the PCCS2 limits, only slightly weaker.} Other basic limits are shown: in the upper shaded region the mass density of black holes exceeds that of all matter in the Universe; the lower grey lines show where $\le 1$ SLAB is expected in a cosmologically relevant volume. The PCCS2 constraints saturate slightly above these incredulity limits only because the 95\% confidence upper limits are compatible with up to $\bar{N} = 3$ SLABs existing \emph{on average} within the volume. \label{fig:SLABLimits}}
\end{figure*}

Figure~\ref{fig:SLABLimits} depicts the resultant limits on SLAB shadows. If we only included limits from the late volume of the universe, over which $\DA$ increases with redshift, then the limits on density steadily increase with SLAB mass (dashed). The most powerful of these late-volume limits comes from the SPT-SZ survey (green; \citep{Everett20}), for which the mean density of SLABs is compatible with the total amount of matter in the Universe when $\MBH \gtrsim 6 \times 10^{14}\ \Msun$. Number density constraints are inversely proportional to the amount of late volume sample by each survey band, which scales as $\omega/F_{\nu}^{3/2}$, where $\omega$ is the solid angle covered and $F_{\nu}$ is the flux density sensitivity. The most interesting results are then for masses of $\sim 10^{16}\ \Msun$, for which $\lesssim 10^{-4}$ of the Universe's mass-energy could be converted into SLABs since $z = 1.6$. The mass of those black holes rivals that of the largest superclusters, however, so it would be straining plausibility for them to form through the accretion of baryonic matter, for example. 

The constraints on growing SLABs follow the late-time limits (dotted), since they are too small at high redshift to benefit from the early volume except when their masses are $\gtrsim 10^{16}\ \Msun$. Since $\MBH \propto (1 + z)$ in this scenario, it can be shown that the flux of a SLAB at redshift $z$ is $|S_{\nu}| \propto \MBH(0)/D_M(z)^2$. But comoving volume is directly proportional to $D_M(z)^3$ \citep{Hogg99}, so the limits are simple power laws extending all the way to the incredulity limit.

SLABs of constant mass, however, loom large in the sky at high redshift, in the early volume. Because of this, there is a threshold mass ($\sim 10^{14} \mathendash 10^{15}\ \Msun$) at which point the shadows of early SLABs becomes visible to the surveys. This immediately opens up an enormous volume, tightening the limits on SLABs by many orders of magnitude, down to $\OmegaBH$ levels of $4 \times 10^{-6}$ (SPT-3G), $2 \times 10^{-7}$ (SPT-SZ), \editOne{$7 \times 10^{-8}$ (VLASS),} and $6 \times 10^{-8}$ (PCCS2). At higher masses, the density constraints become still stronger, although the $\OmegaBH$ limits weaken somewhat because volume gained is insufficient to counter the increased mass of each black hole. Eventually, at $\sim 10^{16} \mathendash 10^{17}\ \Msun$ the limits saturate to a minimum where there can be only $\bar{N}$ SLABs \editOne{on average} in the entire comoving volume in the field of view. For \emph{Planck}'s low-frequency bands, which cover the entire sky, this is the best limit that can be derived from the nonexistence of a SLAB in the entire volume out to the last scattering surface (\editOne{an} ``incredulity limit'' \citep{Carr99}). \editOne{Decreasing the threshold mass significantly is difficult, however, because the area of the shadow scales quadratically with $\MBH$.}

\begin{figure}
\centerline{\includegraphics[width=16cm]{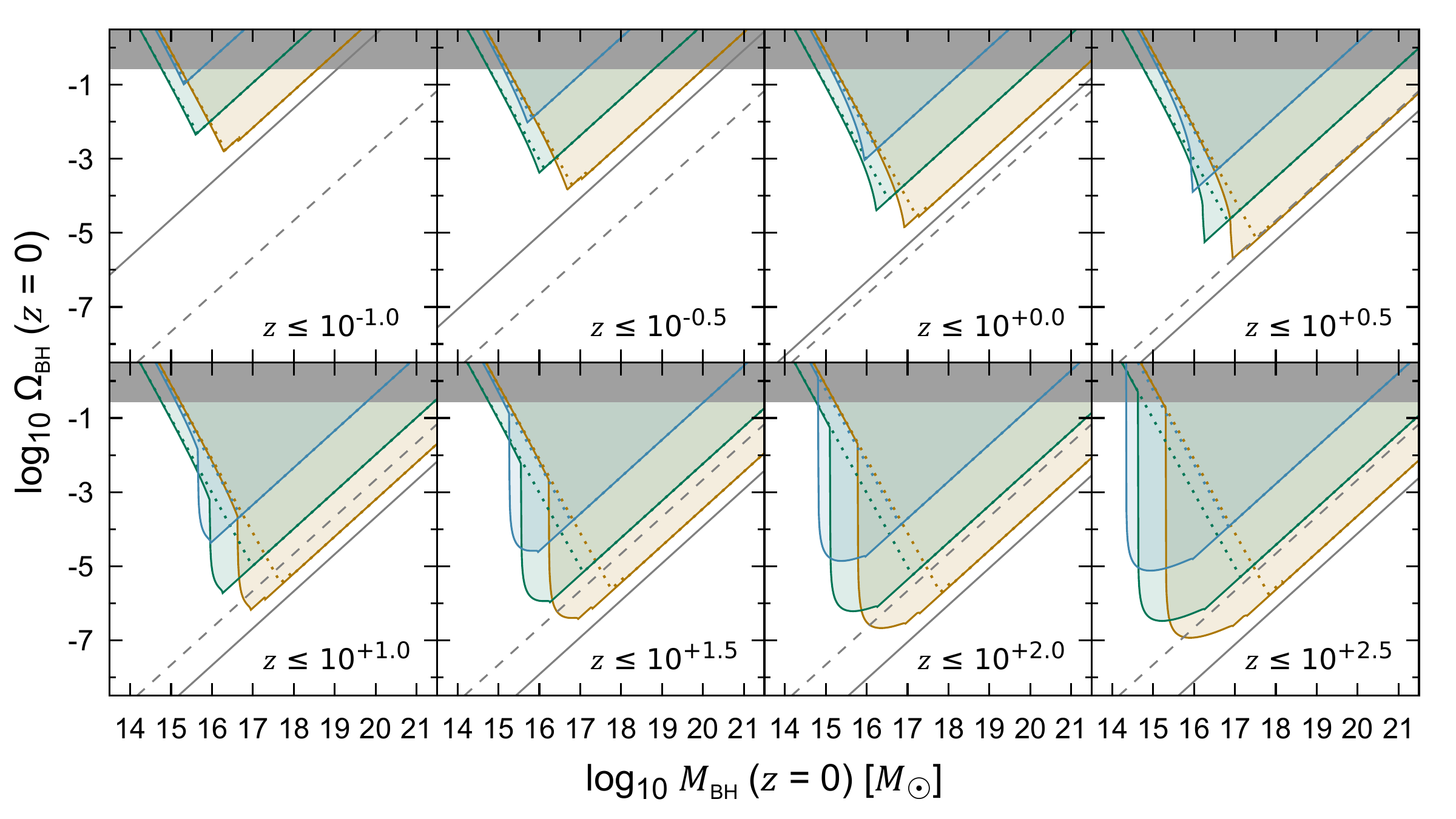}}
\caption{How the shadow limits on SLABs evolve if only the volume within a particular redshift is included. The solid grey line is the incredulity limit for the comoving volume out to that redshift. The colors and other line styles are the same as their counterparts in Figure~\ref{fig:SLABLimits}. \label{fig:LimitEvolution}}
\end{figure}

Carr and K\"uhnel \citep{Carr20} noted that most PBH constraints only set limits on their abundance at a particular epoch, generally either the present or the time before recombination, with little information on their abundance at intermediate redshifts $\sim 10 \mathendash 100$ (with the possible exception of lensing constraints on SLABs at $z \sim 10$ according to \citep{Carr21}). The shadow method is sensitive to these epochs because the largest SLABs can be detected thanks to the reduced angular diameter distances. Indeed most of the comoving volume of the Universe is at these intermediate redshifts, and thus fairly strong limits can be set, as seen in Figure~\ref{fig:LimitEvolution}. The shadow method thus provides constraints on SLABs formed by exotic late-time mechanisms, including SLABs with $\MBH \gtrsim 10^{19}\ \Msun$, in contrast to limits derived from pre-recombination fluctuations in the CMB.

\editOne{By coincidence, the limits that could be derived from the VLASS survey are nearly as powerful as those from \emph{Planck}: the far better sensitivity compensates for the frequency penalty. This suggests that the most powerful radio telescopes operating in the 1-30 GHz range could play a role in setting shadow bounds on SLABs.}

\editOne{\subsection{Comparison with other limits on SLABS}}
\editOne{Several types of limits on SLABs have been proposed in the literature, arising from a variety of possible signatures: (1) $\mu$-distortions to the CMB spectrum resulting when SLABs form from compensated fluctuations \citep{Deng21}, (2) excess isocurvature perturbations in the CMB and baryon acoustic oscillations \citep{Gerlach25}, (3) lensing of the CMB by foreground SLABs \citep{Carr21}, (4) dynamical constraints, in particular, the distortion of the local Hubble flow from nearby SLABs resulting in extreme peculiar velocities \citep{Carr21}, (5) possible radiative signatures from accretion flows or surrounding dark matter haloes \citep{Carr21}, and (6) gravitational wave backgrounds from SLAB formation \citep{DeLuca26}. Some reach the incredulity limit in the accessible cosmological volume. The comparative power depends on the volume in which a SLAB can generate an observable signature and the SLAB mass range where incredulity is reached.}

\editOne{Broadly, the volume probed can be either within the surface of last scattering for signatures created after recombination, outside the surface for signatures detected by inherent fluctuations in the CMB, or both. Black hole shadows against the CMB can only be detected within the last scattering surface. The same applies to lensing, radiative signatures, and detections of gravitational waves by B-mode polarizations. The intrinsic perturbations in the CMB from $\mu$- or isocurvature distortions instead all must be created before recombination. Baryon acoustic oscillations freeze at recombination but are observed in the post-recombination volume, although in practice they currently use tracers like galaxies and the Lyman-$\alpha$ forest practical at relatively low redshifts ($z \lesssim 5$). Finally, directly measured gravitational waves could in principle be emitted before or after recombination. In terms of comoving volume, the post-recombination epoch leads to a nearly optimal incredulity limit: the comoving volume within the last scattering surface is $11,200\,\Gpc^3$ in the \emph{Planck} 2018 cosmology (94\% of the total) with a mere $690\ \Gpc^3$ outside of it (note the small contribution implied in Figure~\ref{fig:ComovingVolume}).}

\editOne{Compare this to the comoving volume that influenced the observed CMB. For any given point on the CMB, the particle horizon distance was $\Delta D_C = D_C(z = \infty) - D_C(z_{\mathrm{CMB}})$, where $D_C$ is the radial comoving distance. Since the CMB records a sphere of comoving radius $D_C (z_{\mathrm{CMB}})$, the comoving volume in causal contact fills a shell with comoving radii between $D_C (z_{\mathrm{CMB}}) - \Delta D_C$ and $D_C (z_{\mathrm{CMB}}) + \Delta D_C$. This shell has slightly less than twice the comoving volume behind the CMB. I find that the comoving volume in contact with the last scattering surface is $1,360\ \Gpc^3$, about 11\% of the total cosmic comoving volume. Of course, there are additional challenges to a direct comparison, in particular, the formation time of SLABs and their growth before recombination, but this fundamental geometric fact means that in principle the post-recombination limits -- shadows, lensing, and gravitational waves -- can more powerfully rule out the largest SLABs.}

\editOne{Just because the shadow limits have access to more comoving volume, and thus a more stringent incredulity limit, that does not mean they are generally stronger than those from the CMB. They have a (typically large) threshold mass, above which the early volume becomes accessible, and only for extreme masses is the entire early post-recombination volume probed. For comparison, the isocurvature limits in ref.\citep{Gerlach25} also reach incredulity, albeit for the pre-recombination volume. The mass range they constrain depends on an ultraviolet cutoff in the perturbation spectrum. Their more conservative limits} establish an upper limit of $\OmegaBH \lesssim 10^{-6} \editOne{\mathendash 10^{-5}}$ over the mass range $\sim 10^{11} \mathendash 10^{19}\ \Msun$. \editOne{A more powerful ``stringent'' isocurvature limit reaches the pre-recombination} incredulity limit for all masses $\gtrsim 10^{14}\ \Msun$. While the shadow constraints \editOne{can be} about an order of magnitude more powerful for $\MBH \gtrsim 10^{14} \editOne{\mathendash 10^{16}}\ \Msun$\editOne{, this only applies to non-accreting SLABs with constant mass. In the accreting SLAB scenarios, the shadow bounds are much weaker. The gravitational wave limits in ref. \citep{DeLuca26} have the potential to reach $\OmegaBH \lesssim 10^{-6}$ for SLAB masses above $\sim 10^{17}\ \Msun$, and to rule out SLABs above a hundred trillion Solar masses as dark matter, although they depend on the poorly known amplitude of the waves.} In addition, ref. \citep{Carr21} noted that the lack of a lensing signature on the CMB from a SLAB could rule out large SLABs ($\gtrsim 10^{16}\ \Msun$) in the redshift range $0.5 \mathendash 10$, although they did not compute it in detail so a more thorough comparison is difficult.

\editOne{Thus, many types of signature yield strong constraints on SLABs, which is perhaps not surprising since their huge masses would necessarily have a disproportionate impact on the Universe. The relative advantages of the shadow method are that it is conceptually simple and has access to the post-recombination comoving volume, larger than that accessible to CMB constraints. It is thus also sensitive to late-forming SLABs, which favors their use at the highest masses. Their major disadvantage is that the sensitivity of CMB experiments still leaves an unconstrained mass window for smaller SLABs, and they are unlikely to push down to $\sim 10^{12}\ \Msun$ in the near future.}

\section{Conclusions}
Carr et al. \citep{Carr21} postulated SLABs with an implicit challenge: what sorts of constraints could we set on them that we simply have not bothered to think of yet? This work presents a very simple, yet powerful limit that comes from a very fundamental trait of black holes: black holes are black. They have shadows, and when they have masses well in excess of a quadrillion Suns, that shadow is visible against the omnipresent CMB with present-day instruments. The limits are especially good for primordial SLABs because objects at high redshift actually appear larger on the sky, as $\DA$ decreases with $z$; most of the comoving volume in the Universe is actually at a relatively small angular diameter distance. The shadows of SLABs with masses of $\sim 10^{17}\ \Msun$ are detectable essentially anywhere out to the surface of last scattering\editOne{, approaching the incredulity limit for the entire observable post-recombination volume.}

The limits are robust if \editOne{no matter accretes onto them}, which is expected if the SLABs form from compensated density fluctuations in the early Universe. Without a compensating void, black holes grow from the ingestion of dark matter and possibly baryonic matter, as in the B85 model \citep[see also][]{Mack07}. This means that the SLABs in the present Universe are much larger than they were in the high redshift regions that dominate the observable comoving volume, weakening the constraints. Infalling baryonic matter may also radiate. The luminosity from this accretion would work against the shadow, but except in certain fine-tuned cases, either the shadow or the accretion flow dominates the net flux, and indeed, radiation from accretion could be much easier to detect than the shadow and imply even stronger limits. Additionally, the baryonic matter could obscure the shadow. However, the conditions in the gas accreting onto a SLAB are highly exotic, and there are basic questions about its physical state -- its temperature, ionization, radiative efficiency, and condensation into stars during the infall -- that remain unexplored.

These\editOne{, and previous limits in the literature,} only scratch the surface of possible constraints on SLABs. Even if we suppose there are no accretion signatures, or we posit an unusual mechanism that results in their formation at $z \sim 0$, nearby SLABs should have a weak lensing signature, their gravitational field distorting the images of background galaxies, as the dark matter in galaxy clusters demonstrates \citep[e.g.,][]{Clowe06}. The potential for SLABs to lens the background CMB has already been noted \citep{Carr21}. The accretion flows of SLABs, if they exist, could also be easily detectable. After all, the accreting halo of a quadrillion solar mass SLAB would be $\sim 10\ \Mpc$ in radius and falling at speeds of hundreds of kilometers per second, with more extreme values for larger SLABs (section~\ref{sec:AccretionFlow}). They could be found in many ways: a kinetic Sunyaev-Zeldovich signature in the CMB, the integrated Sachs-Wolfe effect, cooling radiation as the gas condenses, and possibly an enormous halo of stars formed during infall.

\begin{acknowledgments}
I thank the Breakthrough Listen program for their support. Funding for \emph{Breakthrough Listen} research is sponsored by the Breakthrough Prize Foundation (\url{https://breakthroughprize.org/}). In addition, I acknowledge the use of NASA's Astrophysics Data System and arXiv for this research. I thank the participants of the Dyson Minds 2025 workshop for inspiring me to consider the subject of the largest possible black holes. \editOne{I am grateful to the referee, Eric Agol, and Antonio Iovino for their comments, and to John Beacom for encouragement.}
\end{acknowledgments}

\bibliographystyle{JHEP}
\bibliography{StupendousBHShadows}

\end{document}